\documentclass[dvips]{article}
\usepackage{tabularx}
\usepackage[cp1251]{inputenc}
\usepackage[english]{babel}
\usepackage{cmap}
\usepackage{graphicx}
\usepackage[T2A]{fontenc}

\newcommand{\be}{\begin{equation}}
\newcommand{\ee}{\end{equation}}

\begin{document}
\title{Proposition of direct experiment to study the properties of the neutrino with inverted helicity}
\author{Raul Nakhmanson-Kulish\thanks{Email address:
raoul@iqsystems.ru}\\ \\IQ Systems Ltd.
12, Kravchenko st., 119331 Moscow, Russia\\}
\date{August 3, 2012}
\maketitle

\vfill
\begin{abstract}
Author proposes the idea of direct, affordable in the foreseeable future experiment that could determine whether the neutrino are Dirac or Majorana particles.
\end{abstract}
\vfill
\newpage
\tableofcontents
\newpage

\section{Introduction}

The discovery of the vacuum neutrino oscillations is a clear evidence of the existence of a mass of these particles. Traditionally it was thought that neutrinos are helical particles and always have spin $-1/2$, and an antineutrino $+1/2$, however these views are based on the concept of massless neutrinos. But operating the particle with mass, you can stop it and even send in the opposite direction, maintaining the absolute value of spin, which means the inversion of helicity. Current experimental approaches to the problem of inverted helicity (e.g., search for events of double beta decay) are indirect. In this paper, author would propose the carrying out a direct experiment to obtain and study the properties of the neutrino with inverted helicity.

\section{Neutrino: a Dirac or a Majorana fermion?}

As is known, uncharged fermions having non-zero mass and spin $1/2$ may be subject to one of the two equations: the Dirac equation, similar to that for charged particles such as electron \cite{dirac}, and the Majorana equation, in which the inversion takes helicity particles into antiparticles \cite{majorana}. If the neutrino is a Dirac particle, the inversion of helicity does not change his ways reactions with other particles ($\nu*=\nu$), except that dramatically lowers the cross section of these reactions due to (V-A)-structure of the weak interactions \cite{mohapatra}, that can be detected experimentally. Majorana neutrinos with inverted helicity becomes the antineutrino ($\nu*=\overline{\nu}$) and begins to enter into appropriate reactions, that also can be detected experimentally.

\section{The idea of experiment}

Unfortunately, there are no experimental techniques that can stop the neutrinos and the more fling them back. Therefore, the author considers to be more promising to invert the helicity of neutrino by forming a beam of particles with a velocity faster than a neutrino, as if in a frame associated with beam particles, neutrinos will move back and in the case of Majorana neutrinos behave as antineutrinos. Of course, the reactions proceeding with participation of such neutrinos would be difficult to distinguish from reactions with the antineutrino background. Therefore, the author proposes to use a stable directional flux of low-energy neutrinos and to explore differences in the statistics of reactions involving antineutrinos for different orientations of the probe beam relative to the neutrino flux. If neutrino is a Dirac particle, we should see a marked decreasing of neutrino reactions because inverted Dirac neutrinos are practically sterile
Among all the sources of low-energy neutrinos in the Earth's conditions the most powerful is the Sun. Its use is also convenient to the fact that the daily rotation of the Earth changes the orientation of the probe beam relative to the neutrino flux without any manipulation with the equipment, which is important for the purity of statistics. Another advantage of using solar neutrinos is that thanks to neutrino oscillations we have sufficient  amount of neutrino of any flavor in the flux, and searching for the candidate reaction, we are not limited to any particular flavor.
Thus we come to the following scheme of experiment (view from Northern pole of the Earth):

\begin{figure}[htbp]
	\centering
		\includegraphics[width=0.80\textwidth]{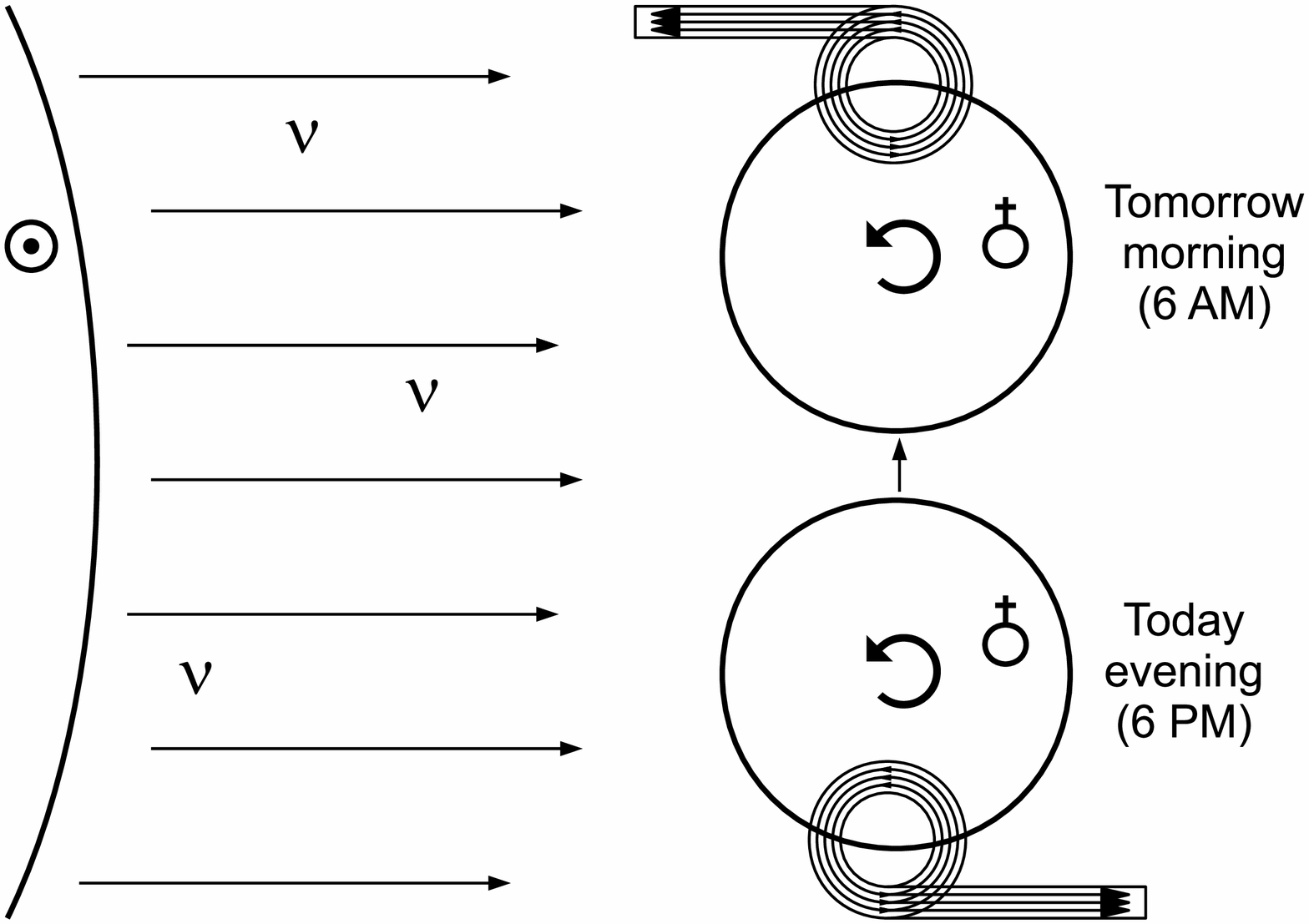}
	\label{fig:exp}
\end{figure}

Beam of high energy particles generated in an accelerator passes some distance by a straight line in the chamber, equipped with detectors of products of neutrino and antineutrino interactions with beam particles. The camera is oriented along the latitude, and in round the clock it has the same direction and the opposite direction relative the solar neutrino flux. Statistics are kept on the number of recorded events depends on time of day.

\section{Neutrino parameters}

According Y.Suzuki \cite{suzuki}, calculated spectrum of solar neutrinos is as follows:
\begin{center}
	\includegraphics[width=0.80\textwidth]{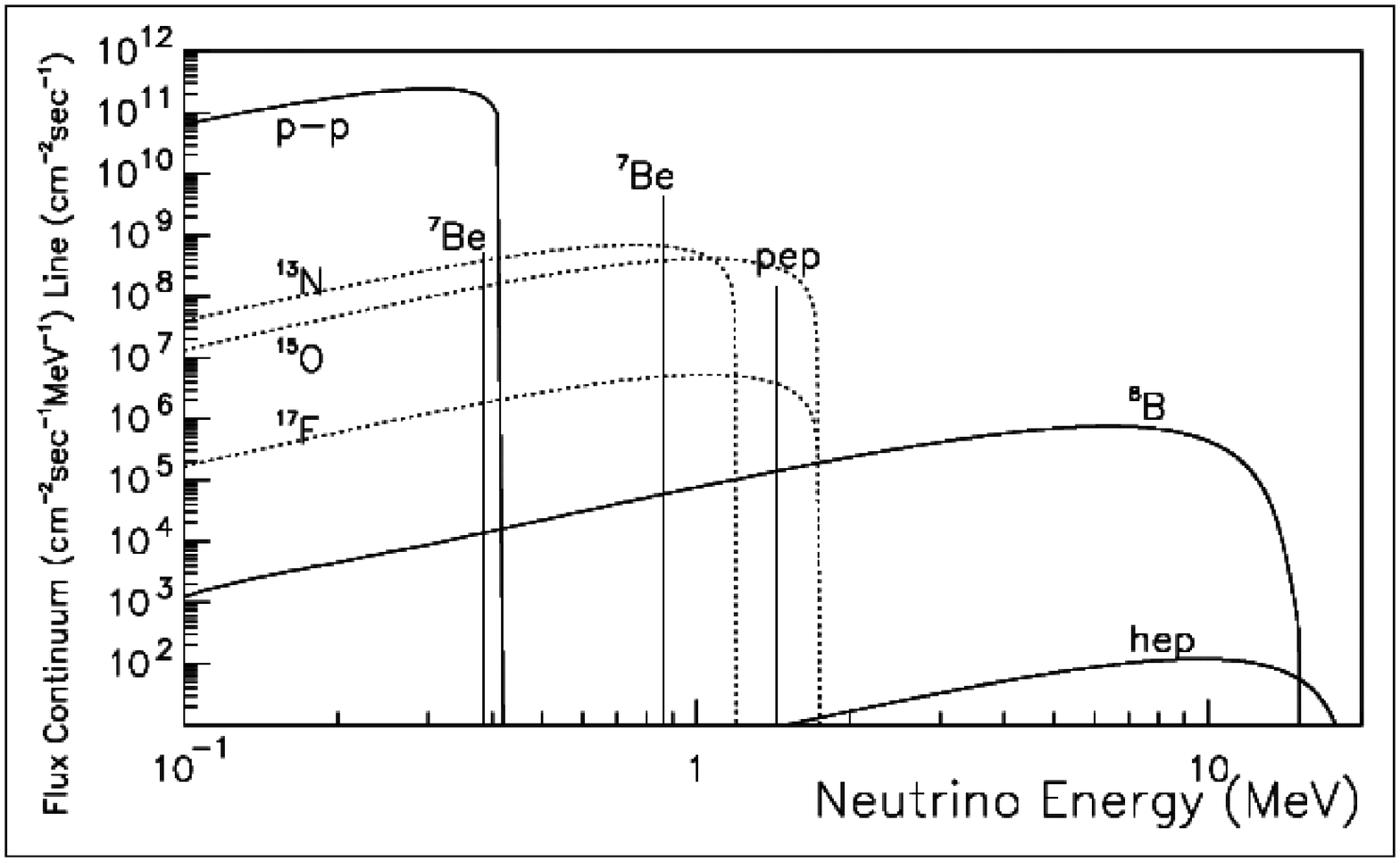}
\end{center}

As can be seen, the proton-proton neutrinos with energy $E \sim 300$ keV dominate in the spectrum, and we will take this value as the basis for kinematic calculations.
Upper limit of neutrino mass sum, according Shaun A. Thomas et al\cite{thomas}, is equal to 0.28 eV, in the same time, the results of C.Amsler et al\cite{amsler} imply that at least one neutrino has a mass not less than 0.04 eV. Let's accept this value as a base estimation of $m_{\nu}$.

\section{Types of reactions}
Reactions, in which inverted neutrinos will come with beam particles, may be divided into endothermic:

\[
\left(\sum m_i\right)_{\rm before} < \left(\sum m_i\right)_{\rm after}
\]

and exothermic:

\[
\left(\sum m_i\right)_{\rm before} > \left(\sum m_i\right)_{\rm after}
\]

We will regard candidate reactions in this order.

\subsection{Kinematics of endothermic reactions}

Using system of units where $c=1$. Let $m_1$ is a mass of probe beam particles, $m_2$ sum of masses of the reaction products. Then the energy adequacy boundary condition of the collision can be written as:

\[
  p_{i\Sigma}p^{i\Sigma} = m_2^2,
\]

where $p_{i\Sigma}$ is a total 4-impulse of particles involved in the collision.

Hence:

\[
  \left( E_1 + E_{\nu} \right)^2 - \left( p_1 + p_{\nu} \right)^2 = m_2^2
\]
\[
  E_1^2 + E_{\nu}^2 + 2 E_1 E_{\nu} - p_1^2 - p_{\nu}^2 - 2 p_1 p_{\nu} = m_2^2
\]
\[
  m_1^2 + m_{\nu}^2 + 2 E_1 E_{\nu} - 2 p_1 p_{\nu} = m_2^2
\]
\[
  2 E_1 E_{\nu} - 2 p_1 p_{\nu} = m_2^2 - m_1^2 - m_{\nu}^2
\]

Let introduce a notation:

\[
  M = \frac{m_2^2 - m_1^2 - m_{\nu}^2}{2}
\]

Hence we have:

\[
  E_1 E_{\nu} - p_1 p_{\nu} = M
\]
\[
  p_1^2 p_{\nu}^2 = \left( E_1 E_{\nu} - M \right)^2
\]
\[
  \left( E_1^2 - m_1^2 \right) \left( E_{\nu}^2 - m_{\nu}^2 \right) = E_1^2 E_{\nu}^2 - 2 E_1 E_{\nu} M + M^2
\]
\[
  m_{\nu}^2 E_1^2 - 2 E_{\nu} M E_1 + \left( m_1^2 E_{\nu}^2 - m_1^2 m_{\nu}^2 + M^2 \right) = 0
\]

Roots of this equation:

\[
  E_1 = \frac{E_{\nu} M}{m_{\nu}^2} \pm \sqrt{ \frac{E_{\nu}^2 M^2}{m_{\nu}^4} - \frac{E_{\nu}^2 m_1^2}{m_{\nu}^2} + \frac{M^2}{m_{\nu}^2} + m_1^2}
\]

The minus sign at the radical can be discarded, because it corresponds to the velocity of probe particle to be lesser than neutrino velocity (or even negative velocity, i.e. probe particle moves towards neutrino), and inversion of helicity doesn't occur. Also note that due to the smallness of the neutrino mass the first term makes the main contribution to expression under the radical, the others for approximate calculations can be neglected.

As a result, we have the minimum necessary energy of probe particle:

\[
  E_1 \approx \frac{2 E_{\nu} M}{m_{\nu}^2} = \frac{ E_{\nu} \left( m_2^2 - m_1^2 - m_{\nu}^2 \right) }{m_{\nu}^2}
\]

Neglecting the neutrino mass in the numerator, finally we can get:

\be \label{minendo}
  E_1 \approx \frac{ E_{\nu} \left( m_2 + m_1 \right) \left( m_2 - m_1 \right) }{m_{\nu}^2}
\ee

\subsection{Endothermic candidate reactions}

From \ref{minendo} we can see that the candidate reaction must have a minimum amount and the minimum difference for the total mass of particles before and after the reaction. From this perspective, calculate the energy required for the following reactions:

\begin{center}
\begin{tabular}{ | c | c | }
  \hline
  \textbf{Reaction} & \textbf{Minimal probe particle energy} \\ \hline
  ${\rm e^{-}} + \overline{\nu}_{\rm e} \: \rightarrow \: \mu^{-} + \overline{\nu}_{\mu}$ & $2 \times 10^{24}$ eV \\ \hline
  ${\rm p} + \overline{\nu}_{\rm e} \: \rightarrow \: {\rm n} + {\rm e^{+}}$ & $6 \times 10^{23}$ eV \\ \hline
  $^{3}{\rm He} + \overline{\nu}_{\rm e} \: \rightarrow \: ^{3}{\rm H} + {\rm e^{+}}$ & $2 \times 10^{22}$ eV \\ \hline
\end{tabular}
\end{center}

Thus, we see that the endothermic reactions require tremendous energies and are unlike to be of interest for our purposes in the foreseeable future.

\subsection{Exothermic reactions}

It is clear that in exothermic reactions involving neutrinos the second particle must be unstable. From this perspective, the project of muon collider is interesting, in the beam of which should go such reaction:
\[
  \mu^{-} + \overline{\nu}_{\mu} \: \rightarrow \: {\rm e^{-}} + \overline{\nu}_{\rm e}
\]

This reaction, like any exothermic, will go with any relative velocity of muon and neutrino, the main point here is that the speed of the muon in the laboratory frame must be higher than the speed of neutrino. Hence we can obtain the minimal required value of the muon energy:
\[
  \frac{E_{\mu}}{m_{\mu}} \geq \frac{E_{\nu}}{m_{\nu}}
\]
\be \label{minexo}
  E_{\mu} \geq E_{\nu} \frac{m_{\mu}}{m_{\nu}} \sim 750 {\rm TeV}
\ee

This value is not so far from the energy of perspective projects of muon colliders ($\sim 100$ TeV)\cite{king}, and considering that a significant portion of solar neutrinos has energies substantially less than 300 keV, we may get the effect to be observed even at $E_{\mu} \sim 100$ TeV.

\section{Conclusions}

Ascertainment of nature of neutrinos with inverted helicity is important problem for elementary particles theory, the way of unoblique experimental solution of which has not yet been suggested. A proposed direct experiment would help to solve this task in the quite foreseeable future. In particular, the possibility of such experiment makes sense to be taken into account just now in the design of powerful muon colliders.

\end{document}